\documentclass{aa}  
\usepackage{graphicx}
\usepackage{natbib}
\usepackage{appendix}

\begin{document}
   \title{On the chemical ladder of esters}

  \subtitle{Detection and formation of ethyl formate in the W51 e2 hot molecular core\thanks{Based on observations carried out with the IRAM 30m Telescope. IRAM is supported by INSU/CNRS (France), MPG (Germany) and IGN (Spain).}}

   \author{V. M. Rivilla\inst{1}, M. T. Beltr\'an\inst{1}, Jes\'us Mart\'in-Pintado\inst{2}, F. Fontani\inst{1}, P. Caselli\inst{3} \and R. Cesaroni\inst{1}     
          }

   \institute{Osservatorio Astrofisico di Arcetri, Largo Enrico Fermi 5, I-50125, Firenze, Italia\\
              \email{rivilla@arcetri.astro.it}
     \and
     Centro de Astrobiolog\'ia (CSIC-INTA), Ctra. de Torrej\'on a Ajalvir km 4, 28850, Torrej\'on de Ardoz, Madrid, Spain \\  
       \and Max-Planck Institute for Extraterrestrial Physics, Giessenbachstrasse 1, 85748, Garching, Germany \\
             }

   \date{Received; accepted}

\titlerunning{Detection and formation of ethyl formate in the W51 e2 hot molecular core}
\authorrunning{Rivilla et al.}

 
  \abstract
   {In the last years, the detection of organic molecules with increasing complexity and potential biological relevance is opening the possibility to understand the formation of the building blocks of life in the interstellar medium.  One of the families of molecules with substantial astrobiological interest are the esters, whose simplest member, methyl formate (CH$_3$OCHO), is rather abundant in star-forming regions. The next step in the chemical complexity of esters is ethyl formate, C$_2$H$_5$OCHO. Despite the increase in sensitivity of current telescopes, the detection of complex molecules with $>$10 atoms such as C$_2$H$_5$OCHO is still a challenge. Only two detections of this species have been reported so far, which strongly limits our understanding of how complex molecules are formed in the interstellar medium. New detections towards additional sources with a wide range of physical conditions are crucial to differentiate between competing chemical models based on dust grain surface and gas-phase chemistry.}
   {We have searched for ethyl formate towards the W51 e2 hot molecular core, one of the most chemically rich sources in the Galaxy and one of the most promising regions to study prebiotic chemistry, especially after the recent discovery of the P$-$O bond, key in the formation of DNA.}
   {We have analyzed a spectral line survey towards the W51 e2 hot molecular core, which covers 44 GHz in the 1, 2 and 3 mm bands, carried out with the IRAM 30m telescope.}  
   {We report the detection of the {{\it trans}} and {{\it gauche}} conformers of ethyl formate. A Local Thermodynamic Equilibrium analysis indicates that the excitation temperature is 78$\pm$10 K and that the two conformers have similar source-averaged column densities of (2.0$\pm$0.3)$\times$10$^{-16}$ cm$^{-2}$ and an abundance of $\sim$10$^{-8}$. We compare for the first time the observed molecular abundances of ethyl formate with different competing chemical models based on grain surface and gas-phase chemistry.}
   {We propose that grain-surface chemistry may have a dominant role in the formation of ethyl formate (and other complex organic molecules) in hot molecular cores, rather than reactions in the gas phase.}
{}
   \keywords{}              

   \maketitle
%

\section{Introduction}

The increase in sensitivity and bandwidth of current radiotelescopes is allowing to detect molecules with increasing number of atoms in interstellar and circumstellar environments. Among the nearly 200 molecules already detected, $\sim$60 are considered complex organic molecules (COMs), .i.e., they contain carbon and have $\geq$6 atoms. Since many of these molecules could play an important role in basic prebiotic chemistry, they are considered the building blocks of life. The study of the formation of COMs are crucial to understand how biochemistry could have emerged in the universe. However, although our knowledge of the chemical complexity of the interstellar medium (ISM) is growing, our understanding about the synthesis of COMs is still very limited. Two general scenarios have been proposed for the formation of COMs: gas-phase chemistry triggered by the thermal evaporation of interstellar ices at high temperatures of $T>$ 100 K was first invoked (e.g., \citealt{millar91}), while dust grain surface chemistry at lower temperatures (10 $<T<$80 K) has been introduced in the last decade (e.g., \citealt{garrod06}). The detection of molecules with increasing complexity in different environments will help us to understand how COMs are formed in the ISM.  

One of the families of COMs that attracts substantial astrobiological interest is that of esters, which are essential for many prebiotic processes. Some kind of esters, glycerides, are the basic constituents of human and animal fats and vegetable oils. Phosphoesters are key to the formation of the backbone of the genetic macromolecules ribonucleic acid (RNA) and deoxyribonucleic acid (DNA).
The simplest representative of the ester family is the 8-atom molecule methyl formate (CH$_3$OCHO, hereafter MF). Interestingly, this molecule is rather abundant in many high-mass and low-mass star-forming regions (e.g., \citealt{brown75}; \citealt{cazaux03}; \citealt{favre11}) and in the Galactic Center (\citealt{requena-torres08}). This makes the interstellar search for more complex esters very promising. The next step in ester complexity is the 11-atom molecule ethyl formate (C$_2$H$_5$OCHO, hereafter EF). Despite the increase in sensitivity of current telescopes, this species has been detected so far only towards two hot molecular cores (SgrB2 N and Orion KL; \citealt{belloche09}, and \citealt{tercero13,tercero15}), which clearly shows the difficulty of detecting molecules with such complexity. SgrB2 N and Orion KL are commonly adopted as templates for astrochemical studies because of their chemical richness. However, they are peculiar sources, and might not be fully representative of the chemical content of typical star-forming regions in our galaxy. The SgrB2 N core (with a mass of $\sim$2.4$\times$10$^{3}$ M$_{\odot}$; \citealt{schmiedeke16}) is located in the Galactic Center, where the physical and chemical conditions are unique, thus an extrapolation to other galactic star-forming regions is not straightforward. Moreover, although the astrochemical study of Orion KL is important due to its small distance (414 pc; \citealt{menten07}), this object cannot be considered a typical galactic hot core because of its relatively low mass ($\sim$10 M$_{\odot}$; \citealt{kurtz00}). Therefore, to better understand how complex organic molecules are formed in star-forming regions, more detections towards other hot cores are needed.

In this regard, the W51 star-forming region is a very promising target. Located at a distance of 5.1 kpc (\citealt{xu09}) in the Sagittarius arm of our Galaxy, harbors two compact radio sources (\citealt{scott78}) known as e1 and e2, which are associated with hot molecular cores (\citealt{zhang97,zhang98}) with a mass $\sim$300 M$_{\odot}$. These cores exhibit a rich chemistry with $\sim$50 molecular species already detected (\citealt{liu01,ikeda01,remijan02,demyk08,kalenskii10}). This chemical richness has encouraged several groups to carry out fruitless attempts to detect very complex prebiotic molecules, such as glycine (NH$_2$CH$_2$COOH, 10 atoms; \citealt{snyder05}) or trans-ethyl methyl ether (C$_2$H$_5$OCH$_3$, 12 atoms; \citealt{carroll15}). \citet{kalenskii10} marginally detected several complex species, being ethylene glycol ((CH$_2$OH)$_2$, 10 atoms) and EF (11 atoms) among them. The recent confirmation of the presence of ethylene glycol towards W51 e2 by \citet{lykke15} opens the possibility to confirm also the detection of EF in this region. It is remarkable that this source is now a main target for astrochemical prebiotic studies, after the first detection in a star-forming region of the P$-$O bond, a key piece of the backbone of the ribonucleic acid (RNA) and deoxyribonucleic acid (DNA) (\citealt{rivilla16a}). For all these reasons, we have searched for EF towards the W51 star-forming region.

EF is one of the largest molecules detected in the ISM but its formation is poorly known.
Its millimeter rotational spectrum has been studied by \citet{demaison84} and \citet{medvedev09}. Only two conformers have been identified in experimental studies: i) the {\it trans} conformer, where the C$-$C$-$O$-$C$=$O chain forms a coplanar zigzag; and ii) the {\it gauche} conformer, with an out-of-plane rotation (to the left or to the right) of the ethyl group (C$_2$H$_5$) around the O$-$C bond.
The ground state of the gauche conformer is 94$\pm$30 K higher in energy (\citealt{riveros67}) than that of the {\it trans} conformer. 
The geometry of the conformers was studied by \citet{peng95}, and is presented in Fig. 1 of \citet{medvedev09}.

In this paper, we use a spectral survey towards the W51 e2 hot molecular core, with a total bandwidth of $\sim$44 GHz, covering the 1, 2 and 3 mm bands, carried out with the IRAM 30m telescope. We report the first detection of both conformers of EF towards the W51 e2 hot core, and discuss the implications of these observations on the formation of this complex species by confronting the observed molecular abundances with the predictions of the competing formation models based on grain-surface and gas-phase chemistry.

\begin{table}
\begin{small}
\caption[]{Summary of IRAM 30m observations towards the W51 e1/e2 hot molecular core used in this work.}
\begin{center}
\begin{tabular}{c c c c}
\hline
 Date & $RA_{\rm J2000}$ &  $DEC_{\rm J2000}$ & Frequency coverage \\
       & 19h 23m  & 14$^{\circ}$ 30$^{\prime}$  & (GHz) \\
\cline{2-3}
\hline       
2012 Apr 25     & 43.90s & 32.0$^{\prime\prime}$  & 88.83-96.60  \\ 
                  &             &           & 104.51-112.29  \\ 
\hline
2012 Dec 13      & 43.90s & 34.8$^{\prime\prime}$ & 99.70-106.30  \\ 
                      &             &            & 238.20-245.98  \\ 
\hline
2015 Dec 9       & 43.90s & 32.0$^{\prime\prime}$ & 134.47-142.25 \\ 
                      &             &            & 150.15-157.90  \\ 
\hline
\hline
\end{tabular}
\end{center}
\label{table-observations}
\end{small}
\end{table}   


\section{Observations}
\label{observations}

We have used data from different observational campaigns performed with the IRAM 30m telescope at Pico Veleta (Spain) on April and December\footnote{The 2012 December data were obtained from \citet{lykke15}.} 2012 and December 2015.
All campaigns used the Eight Mixer Receiver (EMIR) coupled with the Fast Fourier Transform Spectrometer (FTS), which provided a spectral resolution of 0.195 kHz.
The total spectral bandwidth covered is $\sim$44 GHz, in the ranges 88.83$-$96.60, 99.70$-$112.29, 134.47$-$142.25, 150.15$-$157.90 and 238.20$-$245.98 GHz.
%
The spectra were exported from the software package CLASS of GILDAS\footnote{http://www.iram.fr/IRAMFR/GILDAS} to MADCUBAIJ\footnote{Madrid Data Cube Analysis on ImageJ is a software developed in the Center of Astrobiology (Madrid, INTA-CSIC) to visualize and analyze single spectra and datacubes (Mart\'in et al., {\it in prep.}).} (see e.g., \citealt{rivilla16b}), which was used for the line identification and analysis. 

The line intensity of the IRAM 30m spectra was converted to the main beam temperature $T_{\rm mb}$ scale, 
using the efficiencies provided by IRAM\footnote{http://www.iram.es/IRAMES/mainWiki/Iram30mEfficiencies}. The half-power beam widths of the observations can be estimated using the expression $\theta_{\rm beam}$[arcsec]=2460/$\nu(\rm GHz)$, and are in the range 10$\arcsec-$27$\arcsec$.
All the observed spectra were smoothed to a velocity resolution of 1.5 km s$^{-1}$.

The position of the 2012 December observations (19$h$ 23$m$ 43.9$s$, 14$^{\circ}$ 30$^{\prime}$ 34.8$^{\prime\prime}$) is shifted by 2.8$^{\prime\prime}$  in declination with respect to the other two epochs (19$h$ 23$m$ 43.9$s$, 14$^{\circ}$ 30$^{\prime}$ 32.0$^{\prime\prime}$). 
We have checked that this shift in position does not significantly affect the line intensities by comparing the April and December 2012 spectra in the range where they overlap: 104.51$-$106.30 GHz. The line intensities of the December data are only a factor $\sim$1.1 higher than those of the April data, within the expected calibration uncertainties. 


\begin{table}
\caption[]{Unblended or slightly blended transitions of the {\it trans} and {\it gauche} conformers of EF detected towards the W51 e2 hot molecular core.}
\begin{center}
\begin{tabular}{c c c c }
\hline
Frequency &  Transition &  $E_{\rm up}$  & $\int$ T$_{\rm mb}\Delta v$   \\
(GHz) &    & (K) & (K km s$^{-1}$)$^{a}$  \\
\hline\hline
\multicolumn{4}{c}{{\it Trans} conformer} \\
\hline
93.3247  &  17(7,11)$-$16(7,10)     & 76   &  0.13$\pm$0.03  \\
93.3247  &  17(7,10)$-$16(7,9)     &  76  &  0.13$\pm$0.03 \\
93.356781     &  17(6,12)$-$16(6,11)      &  66  &  0.16$\pm$0.04 \\
93.356798  &   17(6,11)$-$16(6,10)    &  66  & 0.16$\pm$0.04  \\
93.41211 &  17(5,13)$-$16(5,12)     &  58  & 0.18$\pm$0.04  \\
93.413115  &  17(5,12)$-$16(5,11)     &  58  & 0.18$\pm$0.04  \\
94.608041  &   17(1,16)$-$16(1,15)    &   42 & 0.26$\pm$0.06  \\
95.313367  &    18(0,18)$-$17(0,17)   &  44  & 0.27$\pm$0.06  \\
95.517215  &   17(2,15)$-$16(2,14)    &  44  & 0.25$\pm$0.06  \\
101.213864  &  18(2,16)$-$17(2,15)     &  49  & 0.30$\pm$0.07  \\
104.49469     &    19(3,17)$-$18(3,16)   &  57  &  0.31$\pm$0.07 \\  
105.23465     &    20(1,20)$-$19(1,19)   &  54  &  0.35$\pm$0.08 \\
105.27195     &    19(1,18)$-$18(1,17)   &  52  &  0.34$\pm$0.08 \\
105.44706     &    19(3,16)$-$18(3,15)   &  57  &  0.32$\pm$0.07 \\
106.89024     &    19(2,17)$-$18(2,16)   &  54  &  0.35$\pm$0.08 \\      
108.55223     &    20(2,19)$-$19(2,18)   &  58  &  0.37$\pm$0.08 \\ 
109.976377      &    20(5,15)$-$19(5,14)   & 73   &  0.30$\pm$0.06 \\
109.982644      &   20(3,18)$-$19(3,17)    &  62  & 0.36$\pm$0.08  \\
110.103357     &  20(4,17)$-$19(4,16)     &  67  & 0.33$\pm$0.07  \\
110.417497     &  21(1,21)$-$20(1,20)     &  59  & 0.40$\pm$0.08  \\
110.544519   &    20(1,19)$-$19(1,18)   &  57  &  0.39$\pm$0.08 \\
136.272898     &   26(1,26)$-$25(1,25)    &  90  & 0.6$\pm$0.1  \\
136.36963     &   26(0,26)$-$25(0,25)    &  89  & 0.6$\pm$0.1  \\
137.275291     &  25(3,23)$-$24(3,22)     &  92  & 0.6$\pm$0.1  \\
137.48096     &    25(6,20)$-$24(6,19)   & 112  &  0.45$\pm$0.07 \\ 
137.482273     &   25(6,19)$-$24(6,18)    &  112  &  0.47$\pm$0.07 \\
137.805525     &  25(4,22)$-$24(4,21)     &  97  & 0.56$\pm$0.09  \\
138.27859     &    25(4,21)$-$24(4,20)   &  97  &  0.56$\pm$0.09 \\ 
141.50245     &    26(1,25)$-$25(1,24)   &  94  &  0.7$\pm$0.1 \\       
151.20362     &    27(2,25)$-$26(2,24)   &  104  & 0.7$\pm$0.1  \\
151.80493     &    29(0,29)$-$28(0,28)   & 111  &  0.7$\pm$0.1 \\ 
153.70219      &    28(11,17)$-$27(11,16)   &  194  & 0.22$\pm$0.03  \\
153.70219      & 28(11,18)$-$27(11,17)       &  194  & 0.22$\pm$0.03  \\
153.83787     &    28(8,20)$-$27(8,19)   & 153  &  0.40$\pm$0.05 \\ 
153.83787     &    28(8,21)$-$27(8,20)   & 153  &  0.40$\pm$0.05 \\
153.931996    &    28(7,22)$-$27(7,21)   &  142  &  0.47$\pm$0.06 \\
153.932149    &    28(7,21)$-$27(7,20)   &  142  & 0.47$\pm$0.06  \\
154.08046     &    28(6,23)$-$27(6,22)   & 133  &  0.54$\pm$0.07 \\  
154.08500     &    28(6,22)$-$27(6,21)   & 133  &  0.54$\pm$0.07 \\ 
154.39306     &    28(5,23)$-$27(5,22)   & 125  & 0.61$\pm$0.09 \\    
155.37695     &    28(4,24)$-$27(4,23)   & 119  & 0.7$\pm$0.1 \\  
156.91332     &    30(1,30)$-$29(1,29)   & 118  & 0.8$\pm$0.1 \\ 
\hline
\multicolumn{4}{c}{{\it Gauche} conformer} \\
\hline
141.56623      &   20(4,17)$-$19(4,16)     & 173   & 0.11$\pm$0.02 \\
156.01357      &   22(10,13)$-$21(10,12)   & 213   & 0.08$\pm$0.01 \\ 
156.01358      &   22(10,12)$-$21(10,11)   & 213   & 0.08$\pm$0.01  \\ 
157.37580      &   22(3,19)$-$21(3,18)     & 187   & 0.14$\pm$0.02  \\
\hline
\end{tabular}
\end{center}
{$^{a}$ From the LTE model fit (see text).} \\
\label{table-clean-transitions}
\end{table}

\begin{figure*}
\centering
\includegraphics[scale=0.5]{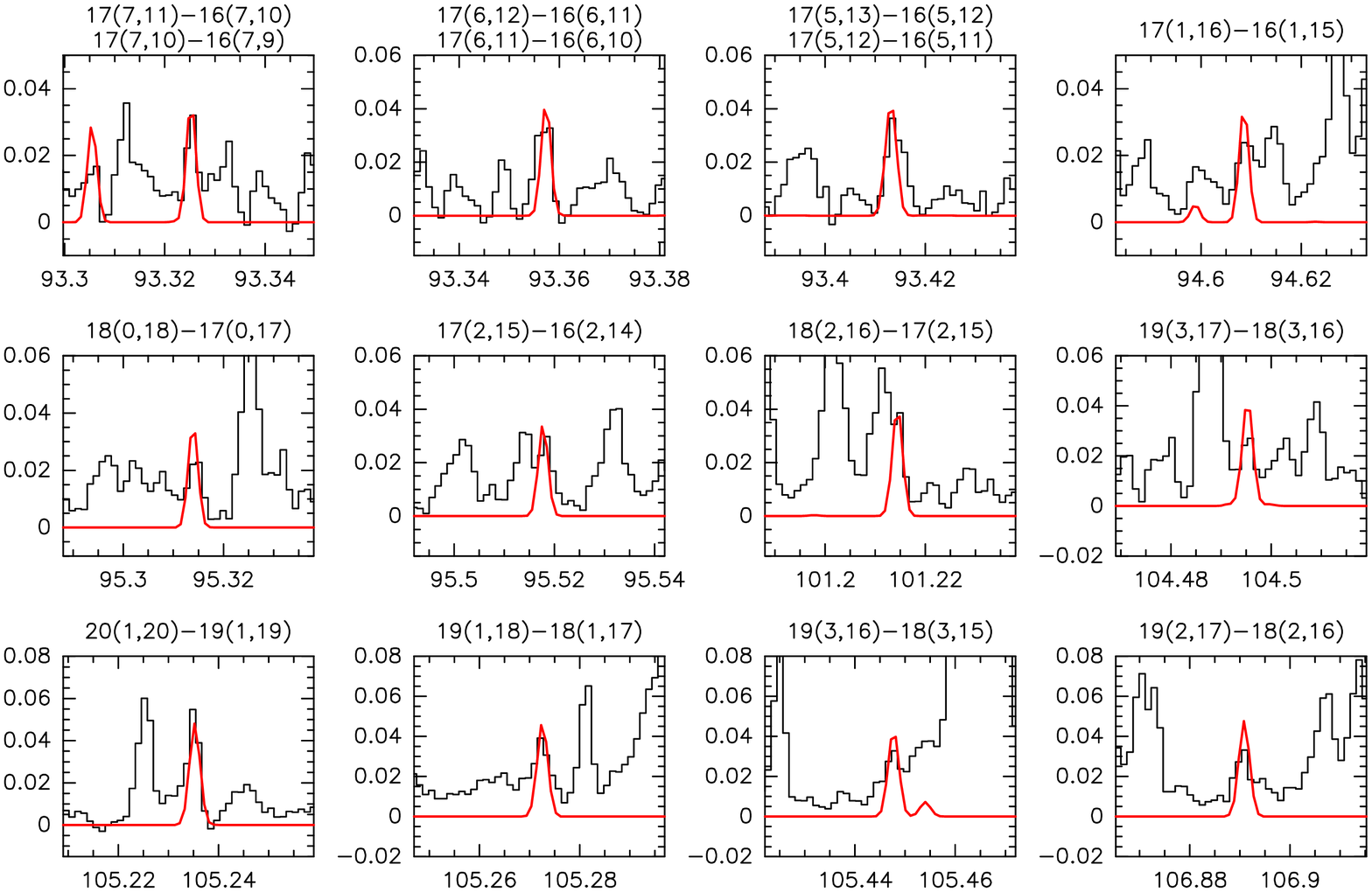}
\vskip3mm
\hspace{-4mm}
\includegraphics[scale=0.5]{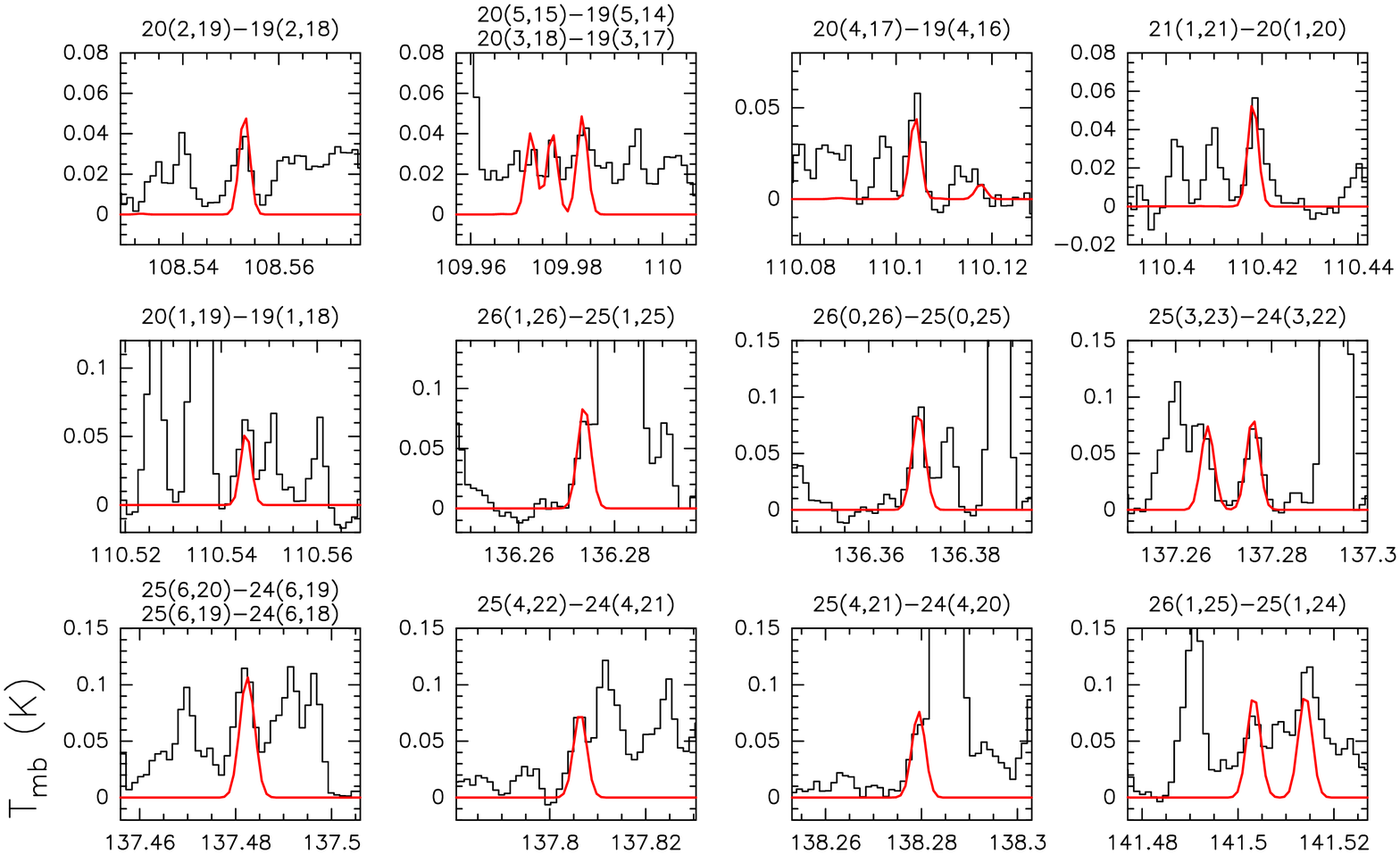}
\vskip3mm
\hspace{2mm}
\includegraphics[scale=0.5]{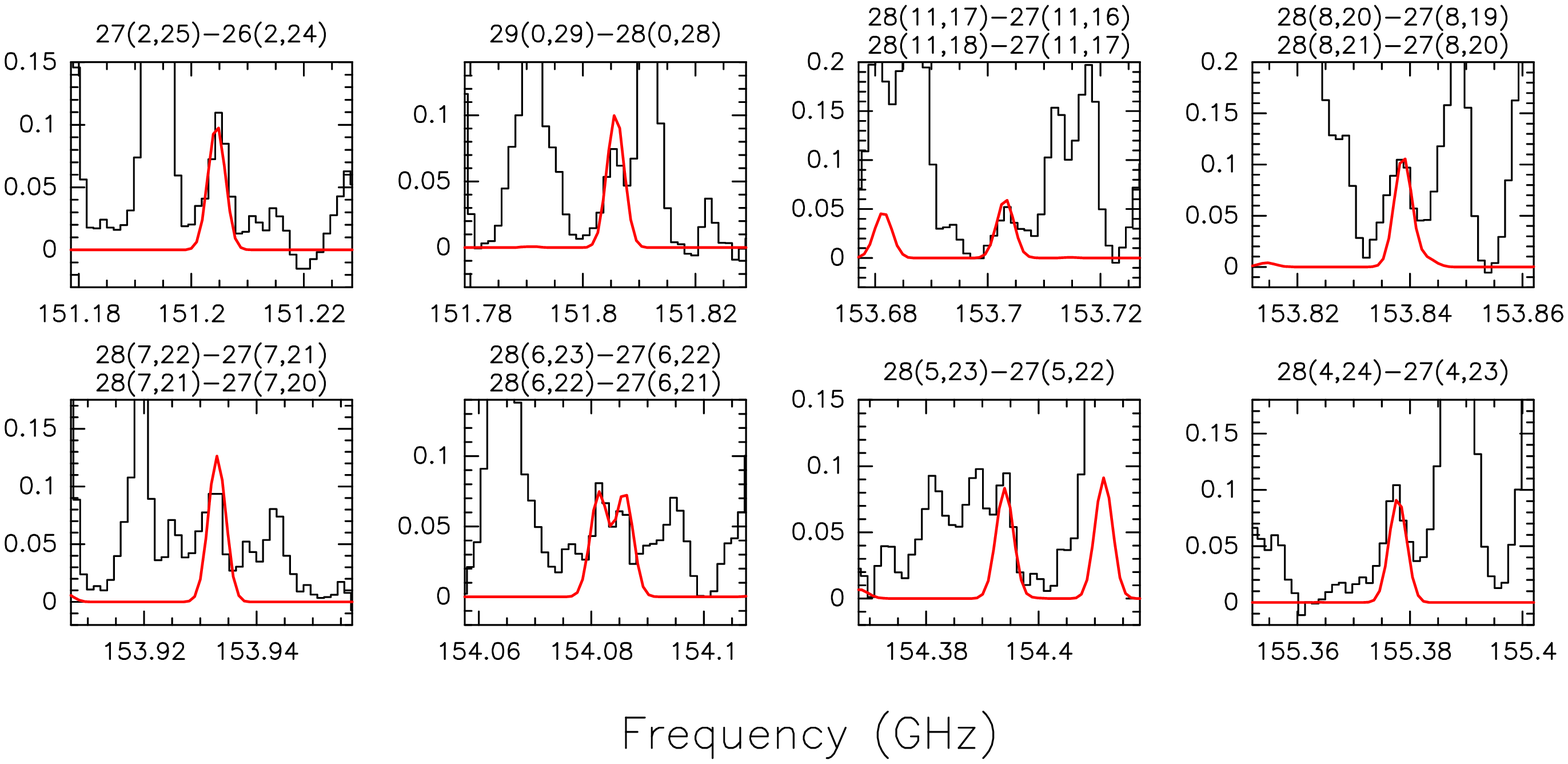}
\caption{Unblended or slightly blended transitions of the {\it trans} conformer of EF towards the W51 e2 hot core (see Table \ref{table-clean-transitions}). We have overplotted in red the LTE synthetic spectrum obtained with MADCUBAIJ assuming a temperature of 78 K, a column density of 2$\times$10$^{16}$ cm$^{-2}$ and a source size of 2.4$\arcsec$.}
\label{fig-clean-transitions}
\end{figure*}

\section{Results}
\label{results}

The observed spectrum contain multiple transitions of both conformers of EF.
We have used MADCUBAIJ to identify the lines and model the expected line profiles for the two conformers of EF assuming LTE conditions. For the analysis we have used the spectroscopic parameters from the Jet Propulsion Laboratory (JPL) molecular catalog\footnote{http://spec.jpl.nasa.gov/}. In Fig. \ref{fig-clean-transitions} we show 42 unblended or slightly blended transitions of the {\it trans} conformer of EF at 2 and 3 mm (see also Table \ref{table-clean-transitions}). The 1 mm spectra exhibit a very high density of molecular lines, preventing the identification of any unblended EF transition.

We have also identified multiple transitions of other molecules that have been proposed to be chemically related to EF: the simplest ester methyl formate (CH$_3$OCHO, hereafter MF), ethanol (C$_2$H$_5$OH, hereafter ET), and formic acid (HCOOH).

The beams of the observations (10$\arcsec-$27$\arcsec$) include two hot molecular cores, W51 e1 and e2, which have different radial velocities of $\sim$59 km s$^{-1}$ and $\sim$55 km s$^{-1}$, respectively (\citealt{zhang98}). To infer which of the two cores is dominating the detected molecular emission we have used the radial velocities of the lines. The velocities of different COMs studied here are in the range 55.2$-$56.4 km s$^{-1}$, and hence they likely arise mainly from e2. In agreement with this interpretation, \citet{kalenskii10} concluded that O$-$bearing species are mainly associated with core e2, while N$-$bearing species are associated with core e1. Therefore, we have considered that the molecular emission of the molecules discussed here mainly arises from W51 e2. 

We have fitted the 42 transitions of the {\it trans} conformer of EF nearly free of contamination (Fig. \ref{fig-clean-transitions} and Table \ref{table-clean-transitions}) to obtain the physical parameters of the emission. MADCUBAIJ takes into account 5 different parameters to model the expected LTE line profiles: source-averaged column density of the molecule ({\it N}), excitation temperature ($T_{\rm ex}$), linewidth ($\Delta$v), peak velocity ({\it v}) and source angular diameter ($\theta_{\rm}$). The procedure used to fit the data was as follows: 
i) we have fixed the size of the molecular emission of W51 e2 to 2.4$\arcsec$ (as derived by \citealt{zhang98}) and applied the beam dilution factor for each transition accordingly;
ii) we have fixed the velocity and the linewidth to the values that reproduce well the observed spectral profile of the unblended lines; and
iii) leaving {\it N} and {\it $T_{\rm ex}$} as free parameters, we used the AUTOFIT tool of MADCUBAIJ, which resolves the radiative transfer equation for a uniform slab and provides the best non-linear least-squared fit using the Levenberg-Marquardt algorithm. 
We have found the best fit for $T_{\rm ex}=$78$\pm$10 K and $N=$(2.0$\pm$0.3)$\times$10$^{16}$ cm$^{-2}$ (Table \ref{table-physical-parameters}). Taking into account that the the H$_2$ column density of W51 e2 is 2$\times$10$^{24}$ cm$^{-2}$ (Ginsburg, {\it priv. comm.}\footnote{The H$_2$ column density has been calculated from the continuum flux at 1 mm, recently observed with ALMA, over a region of 2$"$ size, similar to that of the COMs emission assumed in this work, and assuming a temperature of 100 K, similar to the excitation temperature derived for the COMs (see Table \ref{table-physical-parameters}).}), the abundance of EF is 1$\times$10$^{-8}$.
The line opacities derived from the fits are $\tau\leq$0.07, so we can conclude that the EF emission is optically thin. 
We stress that all the predicted transitions of the {\it trans} conformer of EF with line intensities $>$5$\sigma$ are compatible with the observed spectrum. 


\begin{table*}
\caption[]{Physical parameters of the molecules obtained from the LTE analysis of W51, and comparison with SgrB2 N.}
\begin{center}
\begin{tabular}{c c c c c c c c c}
\hline

Molecule &  \multicolumn{2}{c}{$N$ ($\times$ 10$^{16}$ cm$^{-2}$)}          & \multicolumn{2}{c}{$X$ (10$^{-8}$)} &   \multicolumn{2}{c}{$T_{\rm ex}$ (K)}   & \multicolumn{2}{c}{$\Delta v$ (km s$^{-1}$)}  \\
 & W51 & SgrB2 N & W51 & SgrB2 N & W51 & SgrB2 N & W51 & SgrB2 N \\
\hline\hline
C$_2$H$_5$OCHO (EF)     & 2.0$\pm$0.3 & 5.4 & 1.0  & 0.36 &  78$\pm$10 & 100  & 7.0 & 7.0 \\
C$_2$H$_5$OH  (ET)      & 15.8$\pm$0.4  & 84 & 7.9 & 5.7 &  95$\pm$3  & 100 & 6.9 & 8.0 \\    
CH$_3$OCHO    (MF)      & 50.1$\pm$0.6 & 45 &  25 & 3.5 &  112$\pm$3  & 80 & 6.9 & 7.2 \\
 {\it t}$-$HCOOH   & 1.6$\pm$0.8 & 1.5 &  0.8  & 0.13 & 18$\pm$3   & 70 & 7.1 & 8.0 \\

\hline
\end{tabular}
\end{center}
\label{table-physical-parameters}
\end{table*}

\begin{figure*}
\centering
\includegraphics[scale=0.5]{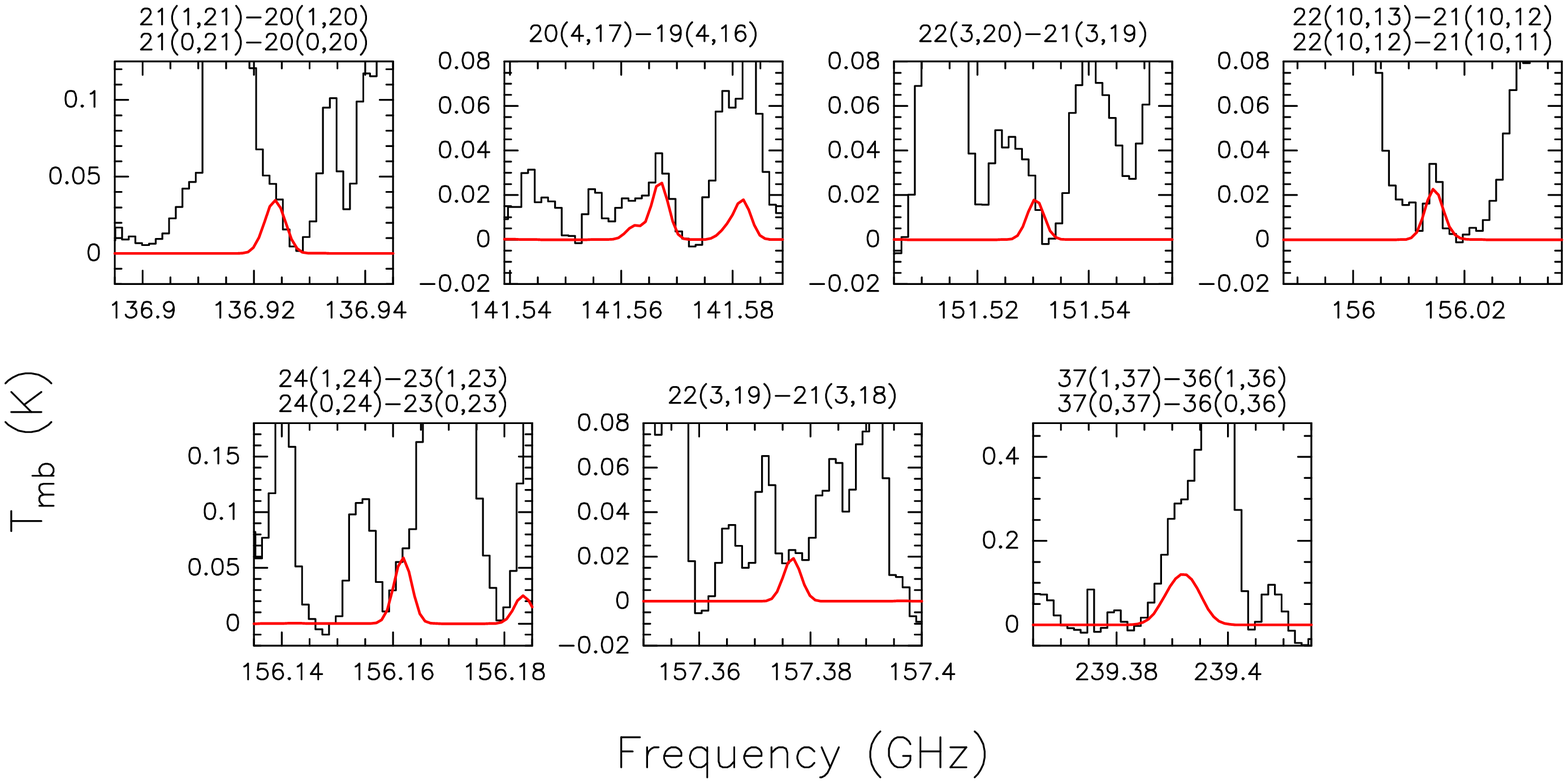}
\caption{Selected transitions of the {\it gauche}-conformer of EF towards the W51 e2 hot core. We have overplotted in red the LTE synthetic spectrum obtained with MADCUBAIJ assuming the same physical parameters as the {\it trans} conformer: $T_{\rm ex}$=78 K, $N$=2.0$\times$10$^{16}$ cm$^{-2}$ and a source size of 2.4$\arcsec$.}
\label{fig-gauche}
\end{figure*}

In the case of the {\it gauche} conformer,
 the predicted LTE line intensities are generally lower than those of the {\it trans} conformer at 2 and 3 mm, which makes its identification more difficult. In Fig. \ref{fig-gauche} we show some selected transitions of the {\it gauche} conformer. The LTE spectrum assuming the same physical parameters used for the {\it trans} conformer reproduces very well the observed spectral features, in particular the transitions at 141.56623 GHz and the two hyperfine transitions at 156.0135\footnote{These transitions were already identified in Orion by \citet{tercero13}; see their Fig. 4.} GHz (see Fig. \ref{fig-gauche} and Table \ref{table-clean-transitions}). All the predicted transitions of the {\it gauche} conformer are compatible with the observed spectra. Therefore, we conclude that the {\it gauche} conformer is also detected in W51 e2, with similar abundance to that of the {\it trans} conformer, as found by \citet{tercero13} in Orion. 
\citet{belloche09} detected the {\it trans} conformer in SgrB2 N and reported only upper limits for the {\it gauche} conformer, although they do not exclude similar abundances for both conformers.

The {\it gauche} conformer has two possible orientations of the ethyl group (C$_2$H$_5$) around the O$-$C bond (see \citealt{medvedev09}). This implies that the relative abundance of both conformers in thermodynamic equilibrium is $N_{\rm g}/N_{\rm t}$=2 e$^{-\Delta E/T_{\rm kin}}$, where {\it g} and {\it t} refer to the {\it gauche} and {\it trans} conformers, respectively, $T_{\rm kin}$ is the kinetic temperature of the gas and $\Delta E$ is the energy difference between the two ground states of the two conformers. The factor of 2 accounts for the two degenerated {\it gauche} conformers. \citet{riveros67} found that this difference is in the range 64$-$124 K. If both conformers are equally abundant, as suggested by the W51 e2 and Orion data, this would imply that the kinetic temperature is in the range $T_{\rm kin}$=93$-$178 K. If we assume that EF is thermalized, then $T_{\rm kin}\simeq T_{\rm ex}$. The excitation temperature derived for EF (78$\pm$10 K; Table \ref{table-physical-parameters}) is slightly lower than the range of $T_{\rm kin}$. This is probably because the transitions of the  {\it trans} conformer (with energies in the range 42$-$194 K) are tracing colder gas than the higher-energy transitions of the {\it gauche} conformer (see Table 2). In any case, we stress that the observed spectra support that both conformers share a similar abundance.

We have also fitted the emission of MF and ET with the same procedure used for EF. Moreover, we have also fitted the emission of the more stable conformer of HCOOH ({\it trans}$-$HCOOH, hereafter t-HCOOH), which has been suggested as a key species to form COMs (\citealt{taquet16}; see Sect. \ref{discussion}). In Appendix B we show some selected unblended transitions of MF, ET and {\it t}$-$HCOOH. In the case of {\it t}$-$HCOOH, since this molecule usually traces gas at lower temperatures (see e.g., \citealt{bisschop07}), one would expect a larger emitting region than that of COMs. For this reason, we have left the source size as a free parameter in the AUTOFIT tool of MADCUBAIJ.
 The derived physical parameters are summarized in Table \ref{table-physical-parameters}.
For {\it t}$-$HCOOH we have found a lower temperature, 18$\pm$3 K, as expected and a source size of 3.2$\arcsec$, only slightly larger than that assumed for the COMs. 
We note that if the size of the emission of HCOOH is larger than this value, then the value of the column density presented in Table \ref{table-physical-parameters} should be considered as an upper limit.

We have also searched for the more excited {\it cis} conformer of HCOOH, which has been proposed to be formed at the surface of cold ices (\citealt{ioppolo11}), but we have not detected any transition. We have obtained an upper limit for the column density of $\sim$10$^{15}$ cm$^{-2}$ (more than an order of magnitude less abundant than {\it t}$-$HCOOH), considering the same temperature, velocity, linewidth and source size than that of {\it t}$-$HCOOH.

\begin{table*}
\caption[]{Gas-phase and grain-surface chemical routes proposed in the literature for the formation of EF.}
\begin{center}
\begin{tabular}{c c c}
\hline
\multicolumn{3}{c}{Gas-phase chemistry}  \\
\hline 
(1)    & C$_2$H$_5$OH (ET) + H$^+$ $\longrightarrow$ C$_2$H$_5$OH$_2^+$ & \citet{charnley95}  \\
\hline
(2a)       & C$_2$H$_5$OH$_2^+$ (ET$^+$) + H$_2$CO  $\longrightarrow$ HC(OH)OC$_2$H$_5^+$ + H$_2$  & \citet{charnley95} \\ 
(2b)    & HCOHOC$_2$H$_5^+$ + $e^{-}$  $\longrightarrow$ C$_2$H$_5$OCHO (EF) + H  & \\ 
\hline
(3a)       & C$_2$H$_5$OH$_2^+$ (ET$^+$) + HCOOH  $\longrightarrow$ HC(OH)OC$_2$H$_5^+$ + H$_2$O & \citet{taquet16}  \\ 
(3b)      & HC(OH)OC$_2$H$_5^+$ + NH$_3$  $\longrightarrow$ C$_2$H$_5$OCHO (EF) + NH$_4^+$ &  \\ 
\hline
\multicolumn{3}{c}{Grain-surface chemistry}  \\
\hline
(4a)  & HCO + 2H $\longrightarrow$ CH$_3$O  & \citet{garrod08} \\
(4b)     &  CH$_3$O + HCO $\longrightarrow$ CH$_3$OCHO (MF) & $\&$ \citet{belloche09} \\
(4c)              & CH$_3$OCHO (MF) + h$\nu$ $\longrightarrow$ CH$_2$OCHO  & \\
(4d)              & CH$_2$OCHO + CH$_3$ $\longrightarrow$ C$_2$H$_5$OCHO (EF) & \\
\hline
(5a)    & HCO + 2H  $\longrightarrow$ CH$_2$OH  & \citet{garrod08}  \\ 
(5b)       & CH$_2$OH + CH$_3$ $\longrightarrow$ C$_2$H$_5$OH (ET) & $\&$ \citet{belloche09} \\ 
(5c)       & C$_2$H$_5$OH (ET) + h$\nu$ $\longrightarrow$ C$_2$H$_5$O & \\ 
(5d)       & C$_2$H$_5$O + HCO $\longrightarrow$ C$_2$H$_5$OCHO (EF)  & \\ 
\hline
\end{tabular}
\end{center}
\label{table-routes}
\end{table*}

\section{Discussion: surface or gas-phase chemistry?}
\label{discussion}

Only very few COMs with $>$10 atoms have been detected in the ISM. In particular, EF (11 atoms) has been reported so far only towards three sources: W51 e2 (this work), SgrB2 N (\citealt{belloche09}) and Orion (\citealt{tercero13,tercero15}). Due to the limited detection of EF in the ISM, it was included only in one chemical model based on grain-surface chemistry (\citealt{garrod08}). This model was considered by \citet{belloche09} to explain the formation of EF in SgrB2 N. Obviously, due to the lack of alternative models, these authors did not discuss other possible formation routes for this species. For Orion, \citet{tercero13} used an energy argument based on gas-phase chemistry to explain the similar abundances of the two EF conformers. However, they did not compare the observed molecular abundances with any chemical model. Only very recently, a new gas-phase chemical network has included the formation of EF (\citealt{taquet16}). This opens the possibility to confront the observed molecular abundances against the predictions of competing models: one based on dust-grain chemistry (\citealt{garrod08}), and the other based on gas-phase chemistry (\citealt{taquet16}).



\subsection{Gas-phase formation}

\begin{figure}
\centering
\includegraphics[scale=0.26]{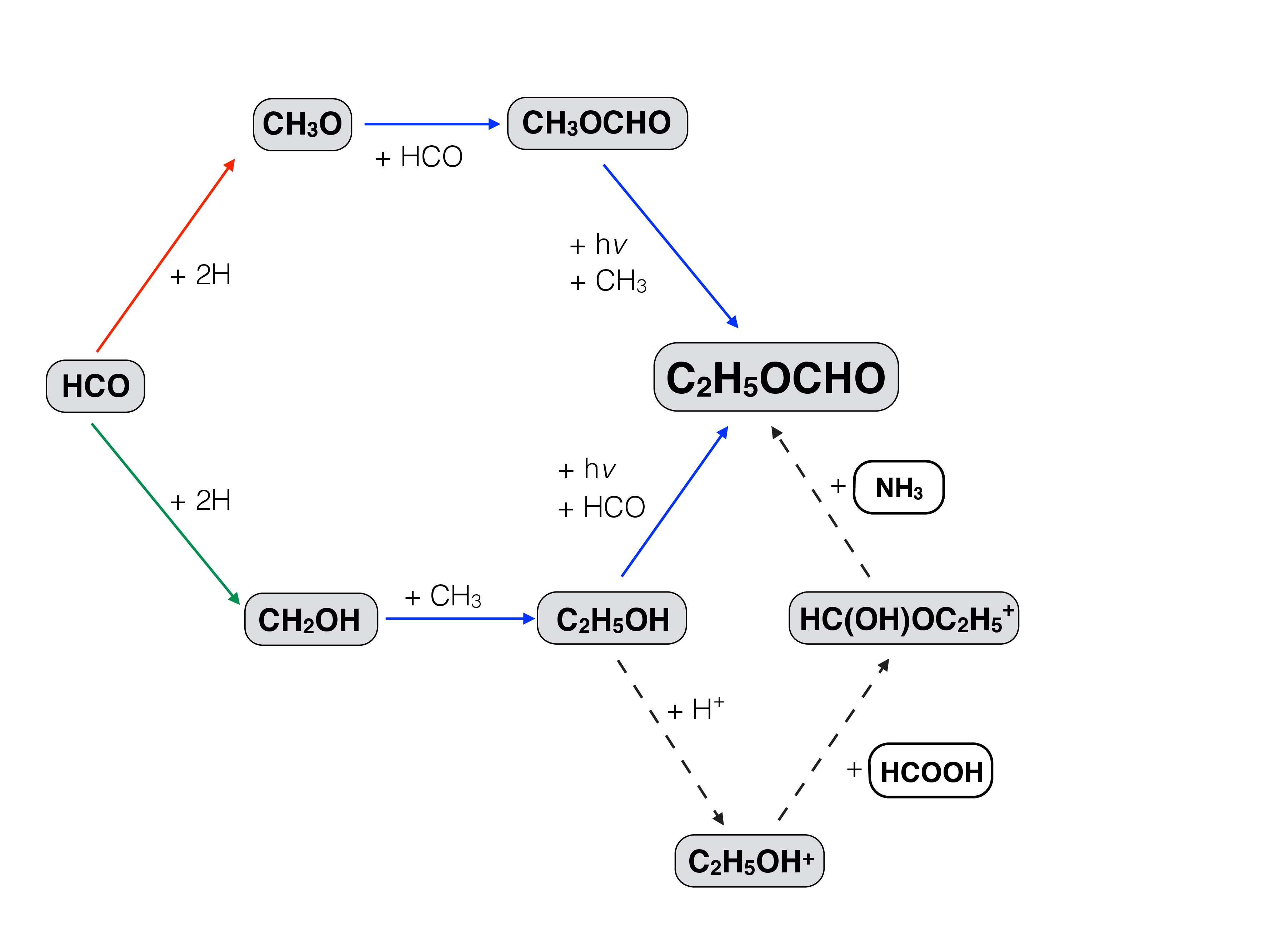}
\caption{Proposed formation routes for ethyl formate. Solid arrows indicate chemical reaction on grain surface, while dashed arrows correspond to gas-phase reactions. The references for the different pathways are: \citet{tielens97,beltran09,woods12} (red arrow); \citet{hama13} (green arrow); \citet{garrod08,belloche09} (blue arrow); and \citet{taquet16} (dashed black arrows).}
\label{fig-routes}
\end{figure}

In the following we will refer to the chemical reactions shown in Table \ref{table-routes} and Fig. \ref{fig-routes}.
\citet{charnley95} proposed a chemical pathway through gas-phase reactions starting with the protonation of ET, followed by a reaction with H$_2$CO and subsequent recombination (reactions 1, 2a and 2b). According to their model, this gas-phase route is able to produce EF abundances of $\sim$10$^{-10}$ in 10$^{4}-$10$^{5}$ yr, comparable to the timescales of hot cores (\citealt{fontani07}). The predicted abundance of EF is however more than 1-2 orders of magnitude lower than that estimated in W51 e2 and SgrB2 N, which suggests that this gas-phase route is not efficient enough to explain the observed abundance of EF.
Moreover, reactions 2a and 2b have not been studied yet in detail in the laboratory. An equivalent gas-phase chemical pathway to form MF, i.e. the protonation of methanol (CH$_3$OH) followed by a reaction with H$_2$CO, has also been proved inefficient by quantum theoretical calculations and laboratory experiments (\citealt{horn04}).
This suggests that the reaction of protonated ET with H$_2$CO (reaction 2a) might also be inefficient to form EF\footnote{We encourage the developement of new laboratory experiments to confirm this.}. Additional works (\citealt{geppert06,vigren10,hamberg10a,hamberg10b}) showed that electron recombination reactions are mostly dissociative, which makes reaction 2b unlikely. 



Very recently, gas-phase chemistry has been reinvoked to explain the formation of COMs. New gas-phase formation routes based on laboratory experiments have been proposed to explain the formation of MF (\citealt{balucani15}). In the case of EF, the new theoretical chemical network by \citet{taquet16} indicates that the combination of protonated ET with HCOOH and NH$_3$ (reactions 3a and 3b) are able to increase the EF abundance by more than one order of magnitude, making this route a promising way to form EF. 
Indeed, this gas-phase model is able to predict an abundance of EF of 10$^{-9}-$10$^{-8}$ (\citealt{taquet16}), similar to that measured (Table \ref{table-physical-parameters}). In this scenario, the presence of enough HCOOH and NH$_3$ is key to allow the formation of EF via the chemical reactions 3a and 3b.  According to this model, to achieve the observed molecular abundances of EF, (0.36$-$1)$\times$10$^{-8}$, the reqired abundances of HCOOH and NH$_3$ are of a few 10$^{-7}$ and $\sim$10$^{-6}$, respectively (Fig. 6 of \citealt{taquet16}).
However, we have estimated that the abundance of HCOOH in W51 e2 is $\sim$10$^{-8}$, while \citet{belloche09} found a value of 1.3$\times$10$^{-9}$ towards SgrB2 N. These abundances are lower than those required for the gas-phase model by a factor $>$50. 

Regarding NH$_3$, \citet{zhang97} derived a column density of $\sim$10$^{17}$ cm$^{-2}$, which implies an abundance of {\bf 5$\times$10$^{-8}$},  which is also a factor 50 lower than that required by the gas-phase model. Therefore, we conclude that these massive star-forming regions do not have high enough abundance of HCOOH and NH$_3$ to form efficiently EF and other COMs via the proposed gas-phase reactions.

\subsection{Grain-surface formation}

The lack of viable gas-phase reactions to form MF and EF (and other COMs) triggered in the past years the development of models based on grain-surface chemistry.   
\citet{garrod08} proposed that EF can be formed from the combination of simpler radicals, such as HCO, CH$_3$O and CH$_3$ (reactions 4 and 5). MF is the primary source of precursor radicals of EF via photodissociation (reaction 4c). Once MF is desorbed from grains, then ET is the main source of precursor radicals (reaction 5c).
The grain-surface model by \citet{garrod08} predicts an abundance of EF of 2.3$\times$10$^{-9}$ (see \citealt{belloche09}), in reasonably good agreement with the range of observed values in W51 e2 and SgrB2 N: (3.6$-$10)$\times$10$^{-9}$.  
%
%
%
%
In addition, unlike the gas-phase model, the HCOOH abundance predicted by the grain-surface model is much closer to the observed values. The [HCOOH/EF] ratio predicted by the surface chemistry models is $\sim$2, in reasonable agreement with the values of $\sim$0.4$-$1 found in W51 e2 and SgrB2 N. The abundances of the COMs (i.e., EF, MF and ET) predicted by the grain-surface model (0.23$-$17)$\times$10$^{-8}$ are also consistent with those observed in W51 e2 and SgrB2 N (Table \ref{table-physical-parameters}). 

Moreover, the excitation temperatures derived for EF in W51, SgrB2 and Orion KL (in the range $\sim$78$-$135 K; this work, \citealt{belloche09} and \citealt{tercero13}) are in good agreement with the temperature associated with the peak gas-phase molecular abundance of the Garrod et al. model (110 K, see Table 15 of \citealt{belloche09}), further supporting the formation of EF on dust grains and subsequent evaporation to the gas phase.

Therefore, we conclude that EF and the other COMs detected in massive star-forming cores are likely formed on dust grains and then desorbed to the gas phase.

\section{Summary and conclusions}

We report on the detection of the two conformers ({\it trans} and {\it gauche}) of ethyl formate towards the W51 e2 hot molecular core. Our LTE line profile analysis indicates that ethyl formate has an excitation temperature of 78$\pm$10 K and that both conformers have the same column density of (2.0$\pm$0.3)$\times$10$^{16}$ cm$^{-2}$, which yields to an abundance of 1.0$\times$10$^{-8}$. 
We have discussed the formation of this complex species in the context of the two available competing chemical models based on dust grain surface and gas-phase chemistry.
We have derived the abundances of ethyl formate, other related complex molecular species and proposed precursors found in star-forming regions, and compared them with the predictions of chemical models. We find that complex organic molecules in hot cores are likely formed on the surface of dust grains and desorbed subsequently to the gas phase at higher temperatures. 

Although the current data support grain-surface formation rather than gas-phase chemistry, it is clear that more detections of EF in a larger sample of astronomical sources, interferometric maps to better estimate the sizes of the emitting cores and thus the molecular abundances, and more laboratory experiments and theoretical models are needed to draw firm conclusions on the relative role of surface and gas-phase chemistry in the formation of EF and chemically related COMs.

\begin{acknowledgements}
This work was partly supported by the Italian Ministero dell'Istruzione, Universit\`a e Ricerca through the grant Progetti Premiali 2012 - iALMA.
V.M.R. sincerely thanks Vianney Taquet for sharing the results of their gas-phase model before its publication, and Adam Ginsburg for the estimates of the W51 e2 hydrogen column density.
JM-P acknowleges partial support by the MINECO under grants AYA2010-2169-C04-01, FIS2012-39162-C06-01, ESP2013-47809-C03-01 and ESP2015-65597-C4-1.
PC acknowledge the financial support of the European Research Council (ERC; project PALs 320620).
%
\end{acknowledgements}


\bibliographystyle{mn2e}
\bibliography{biblio}

\begin{appendices}
\appendix

\section{Spectra of MF, ET and {\it trans}-HCOOH}

\begin{figure*}
\centering
\includegraphics[scale=0.5]{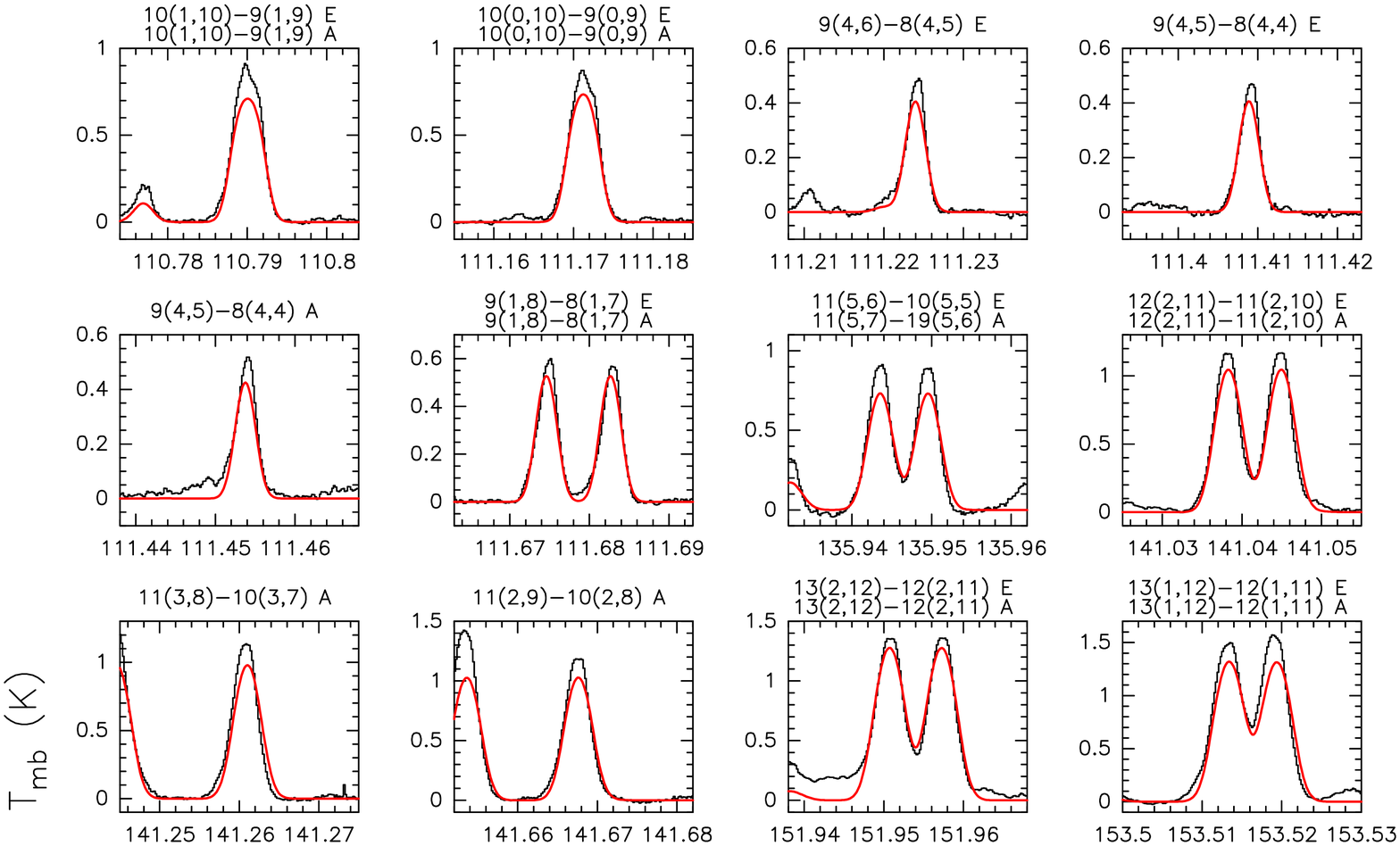}
\vskip3mm
\hspace{3mm}
\includegraphics[scale=0.5]{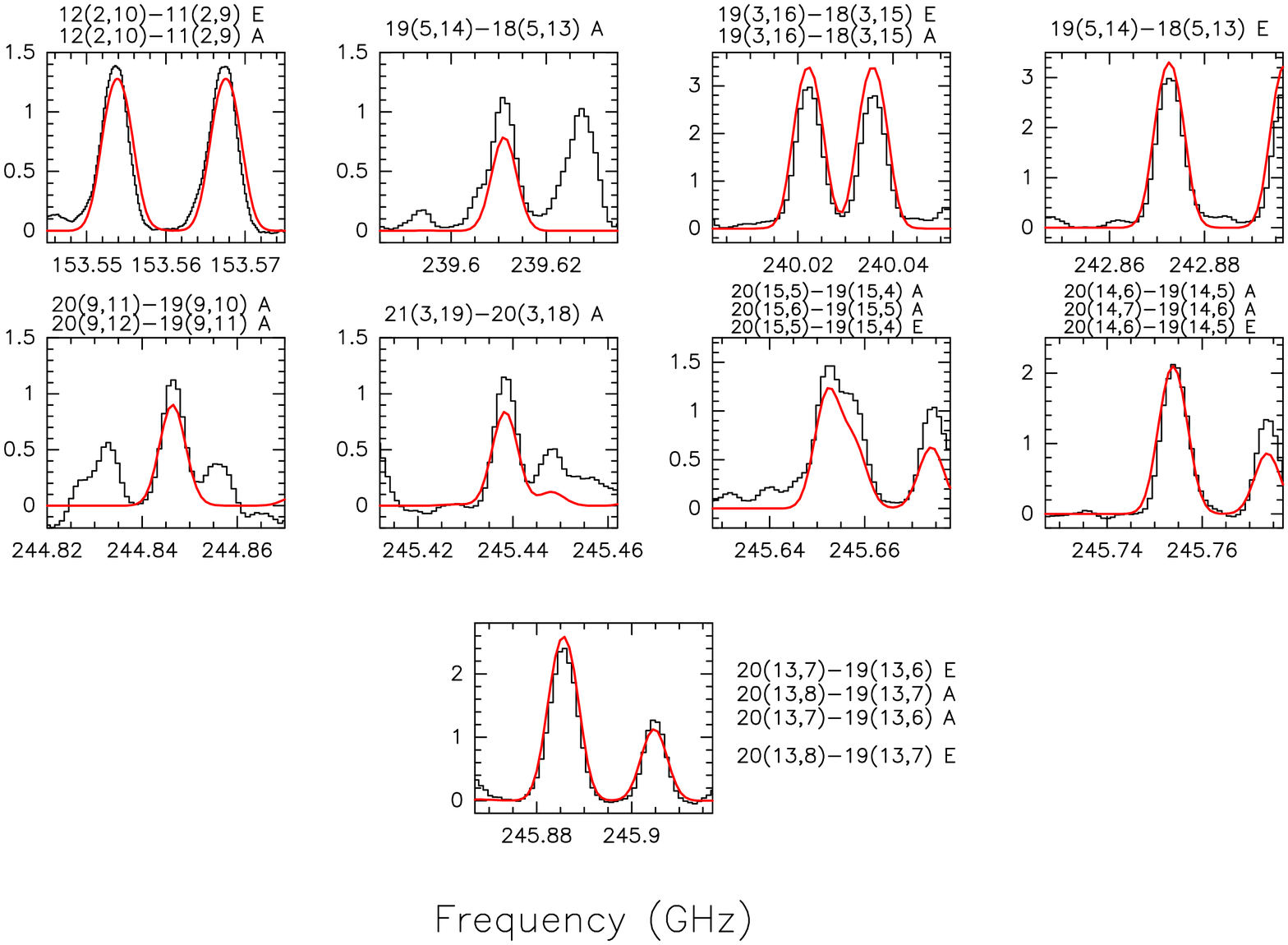}
\caption{Selected unblended transitions of MF (see Table \ref{table-MF}) with LTE fit (red line).}
\label{fig-MF}
\end{figure*}

\begin{figure*}
\centering
\includegraphics[scale=0.5]{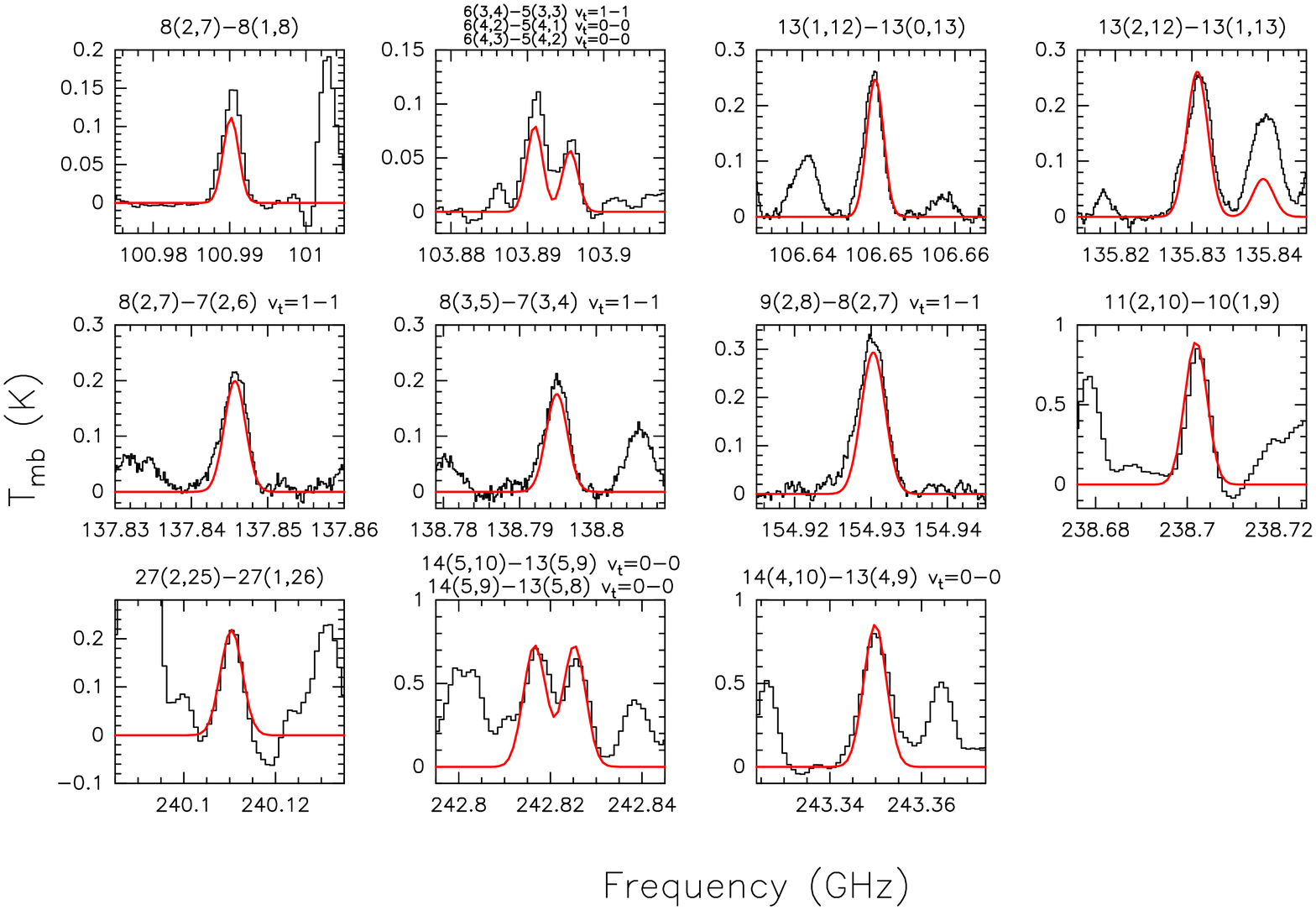}
\caption{Selected unblended transitions of ET (see Table \ref{table-ET}) with LTE fit (red line).}
\label{fig-ET}
\end{figure*}

\begin{figure*}
\centering
\includegraphics[scale=0.5]{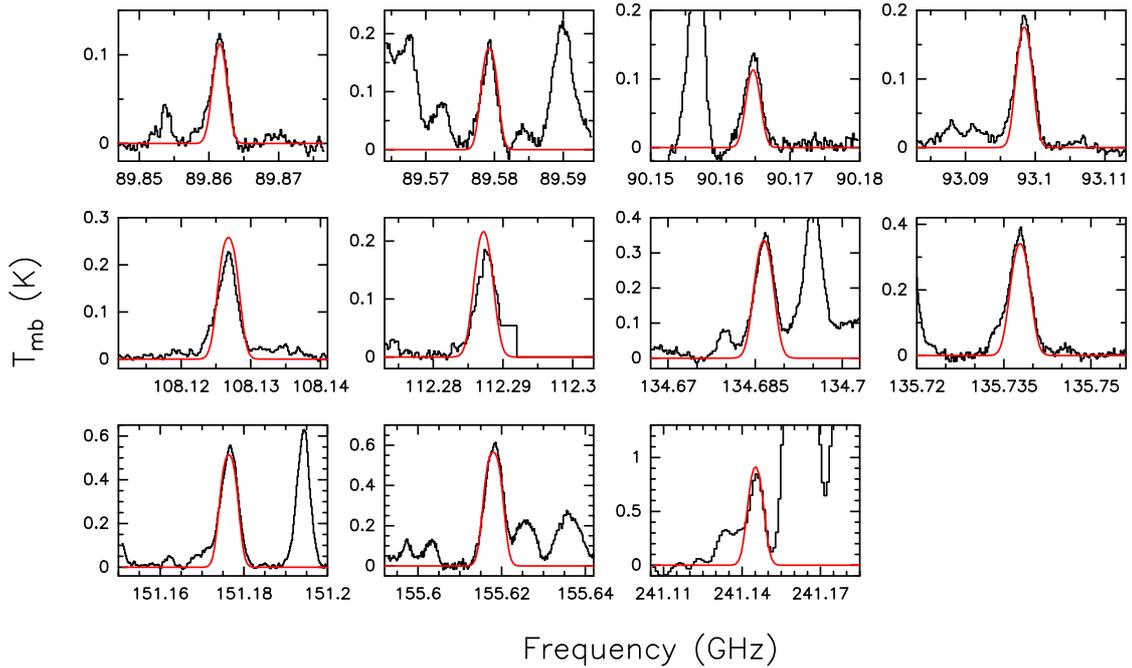}
\caption{Selected unblended transitions of {\it t}$-$HCOOH (see Table \ref{table-app}) with LTE fit (red line).}
\label{fig-HCOOH}
\end{figure*}

We present some selected unblended transitions of MF, ET and {\it t}$-$HCOOH in Figs. \ref{fig-MF}, \ref{fig-ET} and \ref{fig-HCOOH}, and Tables \ref{table-MF}, \ref{table-ET} and \ref{table-app}.

\clearpage


\begin{table}
\caption[]{Selected unblended transitions of MF.}
\begin{center}
\begin{tabular}{c c c c }
\hline
Frequency &  Transition &  $E_{\rm up}$  & $\int$ T$_{\rm mb}\Delta v$   \\
(GHz) &    & (K) & (K km s$^{-1}$)$^{a}$  \\
\hline\hline
110.788664 & v=0, 10(1,10)-9(1,9) E & 30 & 5.0 \\
110.790526 & v=0, 10(1,10)-9(1,9) A & 30 & 5.0 \\
111.169903 & v=0, 10(0,10)-9(0,9) E & 30 & 5.0 \\
111.171634 & v=0, 10(0,10)-9(0,9) A & 30 & 5.0 \\
111.223491 & v=0, 9(4,6)-8(4,5) E & 37 & 3.5 \\
111.408412 & v=0, 9(4,5)-8(4,4) E & 37 & 3.5 \\
111.453300   & v=0, 9(4,5)-8(4,4) A & 37 & 3.7 \\
111.674131 & v=0, 9(1,8)-8(1,7) E & 28 & 4.7 \\
111.682189 & v=0, 9(1,8)-8(1,7) A & 28 & 4.7 \\
135.942980  & v=0, 11(5,7)-10(5,6) A & 56 & 6.4 \\
135.948977 & v=0, 11(5,6)-10(5,5) E & 56 & 6.4 \\
141.037702 & v=0, 12(2,11)-11(2,10) E & 47 & 9.6 \\
141.044354 & v=0, 12(2,11)-11(2,10) A & 47 & 9.6 \\
141.260421 & v=0, 11(3,8)-10(3,7) A & 46 & 8.9 \\
141.667012 & v=0, 11(2,9)-10(2,8) A & 43 & 9.4 \\
151.950079 & v=0, 13(2,12)-12(2,11) E & 55 & 11.9 \\
151.956625 & v=0, 13(2,12)-12(2,11) A & 55 & 11.9 \\
153.512752 & v=0, 13(1,12)-12(1,11) E & 55 & 12.2 \\
153.518739 & v=0, 13(1,12)-12(1,11) A & 55 & 12.2 \\
153.553231 & v=0, 12(2,10)-11(2,9) E & 51 & 11.9 \\
153.566920 & v=0, 12(2,10)-11(2,9) A & 51 & 11.9 \\
239.610154 & v=1, 19(5,14)-18(5,13) A & 317 & 6.2 \\
240.021140 & v=0, 19(3,16)-18(3,15) E & 122 & 32.6 \\
240.034673 & v=0, 19(3,16)-18(3,15) A & 122 & 32.6 \\
242.871569 & v=0, 19(5,14)-18(5,13) E & 130 & 31.0 \\
244.845337 & v=1, 20(9,12)-19(9,11) A & 365 & 3.8 \\
244.845337 & v=1, 20(9,11)-19(9,10) A & 365 & 3.8 \\
245.437310 & v=1, 21(3,19)-20(3,18) A & 327 & 6.9 \\
245.651213 & v=0,  20(15,5)-19(15,4) A & 273 & 5.1 \\
245.651213 & v=0, 20(15,6)-19(15,5) A & 273 & 5.1 \\
245.656780 &  v=0, 20(15,5)-19(15,4) E & 273 & 5.1 \\
245.752266 & v=0, 20(14,6)-19(14,5) A & 254 & 7.1 \\
245.752266 & v=0, 20(14,7)-19(14,6) A & 254 & 7.1 \\
245.754100   & v=0, 20(14,6)-19(14,5) E & 254 & 7.1 \\
245.883179 & v=0, 20(13,7)-19(13,6) E  & 236 &  9.4 \\
245.885243 & v=0, 20(13,8)-19(13,7) A & 236 &  9.4 \\
245.885243 & v=0, 20(13,7)-19(13,6) A & 236 &  9.4 \\
245.903680 &  v=0, 20(13,8)-19(13,7) E & 236 & 9.4 \\
\hline\hline
\end{tabular}
\end{center}
\label{table-MF}
\end{table}


\begin{table}
\caption[]{Selected unblended transitions of ET.}
\begin{center}
\begin{tabular}{c c c c }
\hline
Frequency &  Transition$^{b}$  &  $E_{\rm up}$  & $\int$ T$_{\rm mb}\Delta v$   \\
(GHz) &    & (K) & (K km s$^{-1}$)$^{a}$  \\
\hline\hline
100.990102 &  8(2,7)-8(1,8) &  35 & 0.8 \\
103.890904 &  6(4,3)-5(4,2) $v_{\rm t}$=0$-$0 & 94 &  0.3 \\
103.890904 &  6(4,2)-5(4,1) $v_{\rm t}$=0$-$0 & 94  &  0.3 \\
103.895533 &  6(3,4)-5(3,3) $v_{\rm t}$=1$-$1 &  90 & 0.4 \\
106.649479 & 13(1,12)-13(0,13) &  79 & 1.9 \\
135.830641 & 13(2,12)-13(1,13)  &  81 & 2.0 \\
137.845614 &  8(2,7)-7(2,6) $v_{\rm t}$=1$-$1 &  96 & 1.5 \\
138.794762 &  8(3,5)-7(3,4) $v_{\rm t}$=1$-$1 & 103 & 1.3 \\
154.930188 &  9(2,8)-8(2,7) $v_{\rm t}$=1$-$1 & 104 & 2.2 \\
238.70176 & 11(2,10)-10(1, 9) &  60 & 6.9 \\
240.110238 & 27(2,25)-27(1,26) & 327 & 1.6 \\
242.816446 & 14(5,10)-13(5,9) $v_{\rm t}$=0$-$0 & 175 & 5.5 \\
242.825099 & 14(5,9)-13(5,8) $v_{\rm t}$=0$-$0 & 175 & 5.5 \\
243.349723 & 14(4,10)-13(4,9) $v_{\rm t}$=0$-$0 & 164 & 6.5 \\
\hline\hline
\end{tabular}
\end{center}
$^{b}$ {The transitions of the {\it gauche} state of ethanol are designated with $v_{\rm t}$=0 ({\it gauche +}) and $v_{\rm t}$=1 ({\it gauche -}), while the transions without $v$ number correspond to the {\it trans} state.}
\label{table-ET}
\end{table}


\begin{table}
\caption[]{Selected unblended transitions of {\it t-}HCOOH.}
\begin{center}
\begin{tabular}{c c c c }
\hline
Frequency &  Transition &  $E_{\rm up}$  & $\int$ T$_{\rm mb}\Delta v$   \\
(GHz) &    & (K) & (K km s$^{-1}$)$^{a}$  \\
\hline\hline
 89.57917     & 4(0,4)$-$3(0,3) & 10.8   & 1.7  \\
 89.86148     & 4(2,3)$-$3(2,2) & 23.5   & 1.0  \\
 90.16463     & 4(2,2)$-$3(2,1) & 23.5   & 1.0  \\
 93.09836     & 4(1,3)$-$3(1,2) & 14.4   & 1.7  \\
108.12670     & 5(1,5)$-$4(1,4) & 18.8   &  2.5 \\
112.28712     & 5(2,4)$-$4(2,3) & 28.9   &  1.9 \\
134.68637     & 6(2,5)$-$5(2,4) & 35.4   &  3.1 \\
135.73770     & 6(2,4)$-$5(2,3) & 35.5   &  3.1 \\
151.17628     & 7(1,7)$-$6(1,6) & 32.3   &  5.3 \\
155.61788     & 7(0,7)$-$6(0,6) & 30.0   &  6.0 \\
241.14633     & 11(0,11)$-$10(0,10)     &  70.2  & 8.4  \\
\hline\hline
\end{tabular}
\end{center}
\vspace{5mm}
{$^{a}$ From the LTE model fit (see text).} \\
\label{table-app}
\end{table}

\end{appendices}

\end{document}